\documentclass{llncs}

\usepackage{makeidx}         
\usepackage{graphicx}        
\usepackage{multicol}        
\usepackage[bottom]{footmisc}
\usepackage{url}

\begin{document}

\title{The lifecycle of provenance metadata and its associated challenges and opportunities}
\author{Paolo Missier\inst{1}}
\institute{School of Computing Science \\ Newcastle University \\ Newcastle upon Tyne, UK
\email{Paolo.Missier@ncl.ac.uk}}

\maketitle

\begin{abstract}
This chapter outlines some of the challenges and opportunities associated with adopting provenance principles \cite{cheney_et_al:DR:2012:3507} and standards \cite{Moreau2015a} in a variety of disciplines, including data publication and reuse, and information sciences.\\

\end{abstract}

\begin{figure}
\centering
\includegraphics[width=.8\textwidth]{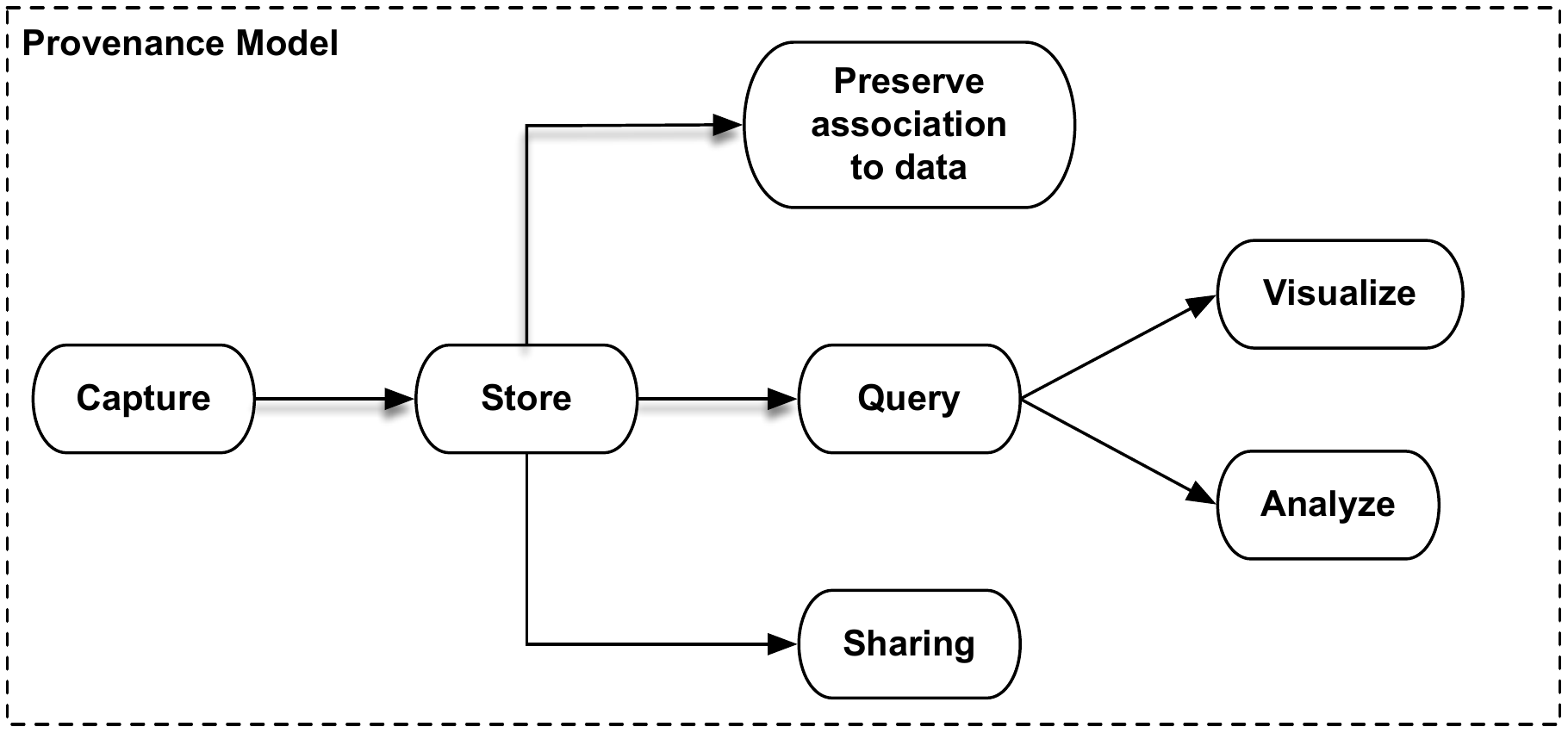}
\label{fig:p-lifecycle}
\caption{Schematic of provenance lifecycle}
\end{figure}

Using provenance in a broad diversity of application areas and disciplines entails a number of challenges, including specialising the generic provenance and domain-agnostic data model, PROV.
This chapter provides a brief overview of these challenges, using the \textit{provenance lifecycle} framework shown in Fig. \ref{fig:p-lifecycle} as a reference.

\section{Provenance definitions and model}

PROV, the Provenance standard, is a family of specifications released in 2013 by the Provenance Working Group, as a contribution to the Semantic Web suite of technologies at the World Wide Web Consortium.
PROV aims to define a \textit{generic} data model for provenance that can be extended, in a principled way, to suit many application areas.
The PROV-DM document \cite{w3c-prov-dm} provides an operational definition of provenance for the community to use and build upon:
\begin{quote}
Provenance is defined as a record that describes the people, institutions, entities, and activities involved in producing, influencing, or delivering a piece of data or a thing. 
\end{quote}
The document goes on to position the definition in the context of Information Management:
\begin{quote}
 The provenance of information is crucial in deciding whether information is to be trusted, how it should be integrated with other diverse information sources, and how to give credit to its originators when reusing it. In an open and inclusive environment such as the Web, where users find information that is often contradictory or questionable, provenance can help those users to make trust judgements.
\end{quote}

\subsection{PROV as a community data model and ontology}

The specifications define a data model and an OWL ontology, along with a number of serializations for representing aspects of provenance. 
The term \textit{provenance}, as understood in these specifications, refers to information about entities, activities, and people involved in producing a piece of data or thing, which can be used to form assessments about its quality, reliability or trustworthiness (PROV-Overview \cite{w3c-prov-overview}). 
The specifications include a combination of W3C \textit{Recommendation} and \textit{Note} documents. 
Recommendation documents include (i) the main PROV data model specification (PROV-DM \cite{w3c-prov-dm}), with an associated set of constraints and inference rules (PROV-CONSTRAINTS \cite{w3c-prov-constraints}); 
(ii) an OWL ontology that allows a mapping of the data model to RDF (PROV-O \cite{w3c-prov-o}), 
and (iii) a notation for PROV with a relational-like syntax, aimed at human consumption (PROV-N \cite{w3c-prov-n}).
All other documents are Notes. These include PROV-XML, which defines a XSD schema for XML serialization \cite{prov-xml-online}.
PROV-AQ, the Provenance Access and Query document \cite{w3c-prov-aq}, which defines a Web-compliant mechanism to associate a dataset to its provenance; PROV-DICTIONARY \cite{prov-dict-online}, for expressing the provenance of data collections defined as sets of key-entity pairs; and PROV-DC \cite{prov-dc-online}, which provides a mapping between PROV-O and Dublin Core Terms.

\subsection{The provenance of PROV}
PROV is the result of a long incubation process within the provenance community. 
The idea of a community-grown data model for describing the provenance of data originated around 2006, when consensus began to emerge on the benefits of having a uniform representation for “data provenance, process documentation, data derivation, and data annotation”, as stated in \cite{Moreau2010a}. 
The First Provenance Challenge \cite{Editorial:Challenge08}  was then launched, to test the hypothesis that heterogeneous systems (mostly in the e-science / cyberinfrastructure space), each individually capable of producing provenance data by observing the execution of data-intensive processes, could successfully exchange such provenance observations with each other, without loss of information. 
The Open Provenance Model (OPM) \cite{Moreau2010a} was proposed as a common data model for the experiment. Other Provenance Challenges followed, to further test the ability of the OPM to support interoperable provenance.

In September, 2009, the W3C Provenance Incubator Group was created. 
Its mission, as stated in the charter \cite{prov-xg-charter}, was to ``provide a state-of-the art understanding and develop a roadmap in the area of provenance for Semantic Web technologies, development, and possible standardization.'' 
W3C Incubator groups produce recommendations on whether a standardization effort is worth undertaking. 
Led by Yolanda Gil at University of Southern California, the group produced its final report in Dec. 2010 \cite{prov-xg-report}. 
The report highlighted the importance of provenance for multiple application domains, outlined typical scenarios that would benefit from a rich provenance description, and summarized the state of the art from the literature, as well as in the Web technology available to support tools that exploit a future standard provenance model.
As a result, the W3C Provenance Working Group was created in 2011, chaired by Luc Moreau (University of Southampton) and Paul Groth (Vrije Universiteit Amsterdam).
The group released its final recommendations for PROV in June, 2013.

\subsection{Other notions of data provenance}
Other formal models of data provenance exist, specifically in the context of database management. 
The provenance of a data item that is returned by a database query, for example, is defined by the semantics of the query itself, and mentions the fragments of the database state that were involved in the query processing \cite{Cheney:2009mb}. An algebraic theory in support of data provenance representation and management has been developed \cite{GreenToddJ.KarvounarakisG.Tannen2007}.
This form of \textit{fine-grained} provenance is often contrasted with \textit{coarse-grained} provenance, which records the input / output derivations that are observed when functions are invoked, typically from within workflows and in the context of scientific data processing \cite{DataEngProvenance_01:}.
Attempts have also been made to reconcile these two views, e.g., when declarative-style queries are embedded within procedural workflow processing \cite{Amsterdamer:2011:PLP:2095686.2095693}.

\section{Embracing provenance: status and opportunities}

As illustrated in Fig. \ref{fig:p-lifecycle}, there are a few key phases in the lifecycle of a provenance document: Production (Capture), persistent storage, Query, Sharing, Association with the underlying data products, and consumption/exploitation (Visualization/Analysis).
The remainder of this short overview will only cover issues concerning Capture, Storage and Query, and Analysis, using the following simple example to illustrate key issues in each of these phases.

In PROV, a provenance document is a set of assertions about the derivations that account for the production of a dataset, including, when available, its attribution.
For example, one can use PROV to formally express the following facts:
\begin{quote}
\small 
Alice took draft v0.1 of paper $P$, made some edits during a certain time interval, and produced a new draft v0.2 of $P$.  \\
In doing so, she used papers $p_1$, $p_2$ as reference.  \\
Alice then delegated Bob to do proofreading of $P$ v0.2, which resulted in a new version v0.3 of $P$.  \\
Alice also published a dataset $D$ as supplementary material to $P$, which she has uploaded to a public data repository, for others to discover and reuse.
\end{quote}
These facts can be expressed formally, using either RDF, XML, or PROV-N, the bespoke near-relational syntax mentioned earlier.

\subsection{Extending PROV}

The PROV Working Group worked hard to ensure that PROV can be extended in a principled way, in order to fit the needs of multiple disciplines where expressing the provenance of data may be important.
Specifically, one can (i) use PROV-O, the PROV OWL ontology, in conjunction with other ontologies, in order to provide rich semantic annotations of data, and (ii) extend PROV-O itself with domain-specific provenance concepts.

As an illustration of (i), in the example above one can semantically characterisze data products as ``papers'' of a certain type, along with the associated activities (editing, proofreading) using a suitable vocabulary, while at the same time characterizing their provenance using an RDF serialisation of the example statements above.
As a reference, in the recent past we have demonstrated this capability in our specification of the \textit{Janus} ontology \cite{Missier2010c}.
In brief, provenance and semantic annnotations serve complimentary roles: the former tells the \textit{history} of a data product, while the latter elucidates its \textit{meaning}.

Regarding extending PROV, one notable example is the ProvONE ontology (formerly known as D-PROV) \cite{Missier2013a}, aimed at capturing at the same time the data dependencies that emerge from observations during data creation (known as \textit{retrospective} provenance), as well as the static structure of the process that is responsible for the generation of the process (known as \textit{prospective} provenance) \cite{5557202}.  
The latter is deliberately missing from PROV, owing to its generality. The D-PROV ontology specifically extends PROV to account for the structure of scientific workflows, a specific type of data-generating process that is important in many e-science applications.

In particular, the latest embodiment of D-PROV, called ProvONE \cite{prov-one}, is currently in production use by the DataONE project (\url{dataone.org}).
DataONE, a large NSF-funded project (2010-2018), is the largest Research Data conservancy project in the USA, with a focus on Earth Observational Data and ecology/climate data in particular. 
With a growing federation that already counts tens of member repositories and hundreds of thousands of science data objects, the DataONE architecture places metadata indexing and management at the cornerstone of its data search and discovery capabilities. ``Searching by provenance'' is a new and unique feature that leverages the ProvONE data model, as well as the automated capture of retrospective provenance whenever R or Matlab (and, soon, Python) scripts that access DataONE science objects are executed.

The ProvONE ontology provides a template for extending PROV, which can be used in a number of other domains, as it illustrates proper use of the PROV extensibility points.

\subsection{Provenance capture}

Provenance is the result of observing a data transformation process in execution, including details of its inputs and outputs, be it a database query or a workflow, including processes carried out by humans or only partially automated.
Key questions concerning the recording (``capturing'') of provenance include (i) what provenance-related events can be observed, (ii) what is the level of detail of these observations, and (iii) how does one deal with multiple, overlapping but inconsistent observations?

Regarding scientific data processing, the ability to record provenance relies entirely on the infrastructure on which the processes are executed. 
An increasing number of tools and systems are being retrofitted with provenance recording capabilities, including the best known workflow management systems \cite{DataEngProvenance_01:,CPE:CPE3035}, and more recently, the Python \cite{murtanoworkflow} and the R languages  \cite{lernercollecting,lerner@ipaw2014} for data analytics.
Two specific instances of provenance capture sub-systems for scientific workflows, that we have actively contributed to,  are \cite{Missier2010a}, for the Taverna workflow management system developed in Manchester to support bioinformatics researchers \cite{tavernaMore,Missier2010b}, and for the eScience Central workflow manager \cite{Hiden2011}.

The case of completely automated processes which run in a centralized environment is, however, the simplest possible scenario.
``Human-in-the-loop'' processes are obviously more problematic, and are limited to capturing human interactions with information systems through a user interface. 
Clearly, solutions in this space are necessarily bespoke, with no known publications reporting specific case studies.

In each of these cases, the observations may be available at a specific level of abstraction, which may or may not be appropriate for the type of downstream analysis requirements (see below). 
These range from fine-grained, high-volume, system-level provenance (ie every file I/O operation in the system) \cite{Macko2011}, to ``coarse-grained'' provenance from workflow executions, where only the inputs and outputs of each workflow block can be observed.

As a consequence of these varying levels of details, it becomes necessary to be able to adjust the quantity of information contained in a provenance document, i.e., by creating \textit{views over provenance} that represent abstractions over provenance. In the example above, we could for instance conflate the editing and proofreading activities into one, high-level ``paper preparation'' activity, and ignore the interim v0.2 of $P$.
Our own work on \textit{provenance abstraction} \cite{Missier2014} builds upon prior research \cite{DBLP:conf/vldb/BitonBD07,springerlink:10.1007/978-3-642-22351-8_13}, reflecting the user need not only to simplify the amount of provenance presented to the used, but also to \textit{obfuscate} provenance in order to preserve its confidentiality.

A further complication in provenance capture, is that the observable processes normally take place on multiple, heterogeneous, autonomous and distributed systems, where the corresponding data is scattered. 
The provenance of an end data product must therefore be \textit{reconstructed} by composing multiple, possibly inconsistent, and incomplete provenance fragments harvested from each of those systems.
This is a relevant but under-studied area of research for provenance, with many potential applications that extend well beyond the realm of e-science. 

\subsection{Storage, Retrieval, and Query}

Storing, indexing, and querying provenance documents requires a data layer not unlike that used to store the underlying data products that the provenance refers to. 
Data provenance that describes the history of large volumes of data is itself bound to have a high volume. 
Furthermore, if one includes in the provenance the intermediate data products that are generated as part of a complex data processing pipeline, it is easy to see that  the size of the provenance documents may vastly exceed that of the data whose history it describes.
Older and recent research has been devoted to studying the trade-offs between storing intermediate data products as part of provenance, which may incur a high storage cost \cite{Woodman2015}, as opposed to partially re-computing the data products (`` smart rerun'' \cite{Cohen-Boulakia:2011:SAR:2034863.2034865}).

Issues of dealing with large-scale provenance were addressed in the \textit{BigProv} international workshop organized in 2013 and co-located with the EDBT conference.
A number of submissions contributed to corroborate the hypothesis that the scalability of provenance management systems is becoming a practical problem if interesting analytics are to be derived from it. Amongst these, a study on reconstructing provenance from log files \cite{Ghoshal2013}.

Provenance documents such as the one in our example are naturally expressed in the form of a graph.
This suggests that graph databases (GDBMS) are suitable for their persistent storage, indexing, and querying.
In our past work we have been experimenting with Neo4J, a new generation GDBMS, in order to study the scalability properties of provenance storage. 
In particular, we have developed ProvGen \cite{Firth2014}, a generator of synthetic provenance graphs of arbitrary size and with topology constraints.
ProvGen is designed to create benchmarks for testing the performance of graph-based provenance data layers. 
It can generate provenance documents with millions of nodes and stores them in a Neo4J database.

At the same time, the standard RDF serialization of PROV, which specifies how provenance documents can be expressed using RDF triples that comply with the PROV ontology (PROV-O), lends itself well to storing provenance graphs in existing RDF triple stores.
However, despite the need for testing provenance data layers at scale, and our own past attempts at soliciting contributions that document scalability of provenance storage and query systems (the ProvBench workshop, co-located with \textit{BigProv} (see above), to the best of our knowledge no official  benchmarks have ever been released.
	
\subsection{Provenance Analytics and novel uses for provenance}

With the broad term ``provenance analytics'' we indicate all forms of consumption and exploitation of provenance corpora, once they have been captured and made available through suitable data engineering solutions, alluded to above.
Relevant questions include: what can we learn from a large body of provenance metadata? what techniques and algorithms can be successfully borrowed from the realm of (Big) Data Analytics, in order ot gain insight into data through its provenance?

Much has been made of provenance as a key form of metadata to help understanding the quality of data as well as its trustworthiness. 
A whole special issue of the ACM Journal of Data and Information Quality \cite{Missier:2015:EDI:2698232.2692312}, has been devoted to the topic \cite{Missier:2015:2698232}.
Despite several  high quality submissions, however, more research is needed to fully elucidate the connection between data provenance and quality.

Many other opportunities are worth exploring that exploit provenance corpora in several domains.
One line of research still in its infancy, concerns using provenance to ascribe \textit{transitive credit} \cite{Katz2014} to scientists and other contributors who publish their datasets in public data repositories, for others to reuse. 
Data publication is a rapidly growing area of Open Science, which is based upon the assumption that scientistis will spontaneously make their datasets public, as long as due credit is given to them through community mechanisms.   
Unfortunately, these mechanisms are still quite primitive, limited as they are to counting the number of citations to datasets, as they are found in paper publications (see for instance the \textit{Making Data Count} project \cite{Kratz2015a}). 
Instead, transitive credit pushes this embrionic notion of ``credit for data'' much further, as it leverages provenance to take into account multiple generations of data derivation and reuse.

Other disciplines farther away from computing and science will benefit from properly collected provenance, wherever providing accountability of a process execution is important. 
One example amongst many concerns \textit{food safety}, where traceability of lots of food along a supply chain is critical to ensuring compliance with quality standards and proper handling, and to answer questions in case of accidents involving consumption of unsafe food.

\subsection{Three key challenges for practical usability of provenance data}

To conclude this overview, three areas when more research is needed in order to make provenance usable in practice are worth mentioning.

\paragraph{Incomplete and uncertain provenance.} Generation and usage of data naturally occurs in many different ways through multiple, autonomous information systems. As a consequence, the provenance of such data is also naturally \textit{fragmented and incomplete}. 
One major problem in provenance research is how to reconstruct a complete ``big picture'' out of such fragments. 
We are currently addressing this foundational problem in the specific setting of Open Research Data reuse, as this is a key issue when establishing transitive credit as mentioned above.	

\paragraph{Trusted provenance. } A second issue concerns accountability of the provenance documents themselves. 
To the extent that provenance documents are considered as a form of evidence for the underlying data, it is necessary to ensure that the provenance itself can be trusted not to have been tampered with.
Using provenance traces in, say, a court of law, requires strong non-repudiability and integrity guarantees, which can only be provided by a trusted computing infrastructure \cite{mitchell2005trusted,Lyle:2010:TCP:1855795.1855796}.
The notion of tamper-proof (or rather, tamper-evident) provenance has been touched upon in the past \cite{springerlink:10.1007/978-3-642-04219-5_2}, but more research is needed as this  clearly conflicts with the notion of provenance abstraction through views, alluded to above, namely when generating views involves \textit{redacting} the provenance document itself \cite{Cadenhead:2011:TPU:1998441.1998456}.

\paragraph{Provenance to help the reproducibility of scientific processes.}

Lastly, we mention a long-standing promise on which provenance studies have largely yet to deliver.
Much has been said (and there is no scope for a full survey here) of the role of provenance to support reproducible science, since the connection between reproducibility and provenance was first made back in 2008 \cite{DBLP:conf/sigmod/DavidsonF08}. 

Reproducibility is a known problem for a large number of scientific processes of the past, which are often encoded as a loose collection of scripts with external dependencies on ever-changing libraries, services, and databases.
Practical solutions where provenance is used to ensure that these processes are reproducible are not readily available, however.
In the recent past, we have addressed one aspect of this problem, namely by showing that provenance traces can be used to explain the differences between two sets of results that are obtained from the executions of two versions of a process \cite{CPE:CPE3035}, the latest being a reproduction of the original.
Much remains to be done, however, to clearly prove the role of provenance data in data-driven, reproducible science.


\newcommand{\etalchar}[1]{$^{#1}$}

\end{document}